\documentclass[12pt]{article}

\usepackage{graphicx}

\begin{document}


\title{\bf
Generalized Jacobi Elliptic One-Monopole - Type A\footnote{to be submitted for publication}}

\author{
{\bf Rosy Teh\footnote{e-mail: rosyteh@usm.my}, Khai-Ming Wong and Kok-Geng Lim}\\
{\normalsize School of Physics, Universiti Sains Malaysia}\\
{\normalsize 11800 USM Penang, Malaysia}}

\date{February 2010}
\maketitle

\begin{abstract}
We present new classical generalized one-monopole solution of the SU(2) Yang-Mills-Higgs theory with the Higgs field in the adjoint representation. We show that this generalized solution with $\theta$-winding number $m=1$ and $\phi$-winding number $n=1$ is an axially symmetric Jacobi elliptic generalization of the 't Hooft-Polyakov one-monopole. We construct this axially symmetric one-monopole solution by generalizing the large distance asymptotic solution of the 't Hooft-Polyakov one-monopole to the Jacobi elliptic functions and solving the second order equations of motion numerically when the Higgs potential is vanishing and non vanishing. These solutions are regular non-BPS finite energy solutions. 
\end{abstract}



\section{Introduction}
The 't Hooft-Polyakov one-monopole solution of the SU(2) Yang-Mills-Higgs (YMH) field theory with nonzero Higgs mass and self-interaction is the first monopole solution to possess finite energy \cite{kn:1}. This numerical, spherically symmetric non-BPS one monopole solution is invariant under a U(1) subgroup of the local SU(2) gauge group. All known finite energy solutions of the YMH field theory with a unit magnetic charge are spherically symmetric \cite{kn:1}-\cite{kn:2}. All other finite energy monopole configurations with magnetic charges greater than unity do not possess spherical symmetry \cite{kn:3} as it has been shown that these solutions cannot possess spherical symmetry when energy is finite \cite{kn:4}. 

Exact solutions of the YMH equations only exist in the Bogomol'nyi-Prasad-Sommerfield (BPS) limit when the Higgs potential vanishes or when the Higgs field self-interaction constant $\lambda = 0$. Outside the BPS limit and when $\lambda \not= 0$, only numerical solutions exist. 
The monopole-antimonopole pair (MAP), monopole-antimonopole chain (MAC) and vortex rings configurations of Ref. \cite{kn:5} are non-Bogomol'nyi solutions and they are numerical solutions even in the limit of vanishing Higgs potential. 

Here, we present new classical generalized Jacobi elliptic axially symmetric one-monopole solutions. These new monopoles are regular solutions of the SU(2) YMH theory with finite energies. Similar to the monopole solutions of Ref. \cite{kn:5}, they are non-Bogomol'nyi solutions and are numerical solutions even in the limit of vanishing Higgs potential.


\section{The SU(2) Yang-Mills-Higgs Theory}
The SU(2) YMH Lagrangian in 3+1 dimensions with non vanishing Higgs potential is given by
\begin{equation}
{\cal L} = -\frac{1}{4}F^a_{\mu\nu} F^{a\mu\nu} - \frac{1}{2}D^\mu \Phi^a D_\mu \Phi^a - \frac{1}{4}\lambda(\Phi^a\Phi^a - \frac{\mu^2}{\lambda})^2. 
\label{eq.1}
\end{equation}

\noindent Here the Higgs field mass is $\mu$ and the strength of the Higgs potential is $\lambda$ which are constants. The vacuum expectation value of the Higgs field is $\xi=\mu/\sqrt{\lambda}$. The Lagrangian (\ref{eq.1}) is gauge invariant under the set of independent local SU(2) transformations at each space-time point.
The covariant derivative of the Higgs field and the gauge field strength tensor are given respectively by 
\begin{eqnarray}
D_{\mu}\Phi^{a} &=& \partial_{\mu} \Phi^{a} + g\epsilon^{abc} A^{b}_{\mu}\Phi^{c},\nonumber\\
F^a_{\mu\nu} &=& \partial_{\mu}A^a_\nu - \partial_{\nu}A^a_\mu + g\epsilon^{abc}A^b_{\mu}A^c_\nu.
\label{eq.2}
\end{eqnarray}
Since the gauge field coupling constant $g$ can be scaled away, we can set $g$ to one without any loss of generality. The metric used is $g_{\mu\nu} = (- + + +)$. The SU(2) internal group indices $a, b, c$ run from 1 to 3 and the spatial indices are $\mu, \nu, \alpha = 0, 1, 2$, and $3$ in Minkowski space.

The equations of motion that follow from the Lagrangian (\ref{eq.1}) are
\begin{eqnarray}
D^{\mu}F^a_{\mu\nu} &=& \partial^{\mu}F^a_{\mu\nu} + \epsilon^{abc}A^{b\mu}F^c_{\mu\nu} = \epsilon^{abc}\Phi^{b}D_{\nu}\Phi^c,\nonumber\\
D^{\mu}D_{\mu}\Phi^a &=& \lambda\Phi^a(\Phi^{b}\Phi^{b} - \frac{\mu^2}{\lambda}).
\label{eq.3}
\end{eqnarray}
Non-BPS solutions to the YMH theory are obtained by solving the second order differential equations of motion (\ref{eq.3}), whereas BPS solutions can be obtained more easily by solving the first order Bogomol'nyi equations,
\begin{equation}
B^a_i \pm D_i\Phi^a = 0.
\label{eq.4}
\end{equation}
The static energy for the system is
\begin{eqnarray}
{\cal E} = \int d^3 x~\frac{1}{2}(B_{i}^{a}B_{i}^{a}+D_{i}\Phi^{a}D_{i}\Phi^{a})+\frac{\lambda}{4}\left(\Phi^{a}\Phi^{a}-\frac{\mu^2}{\lambda}\right)^2,
\label{eq.5}
\end{eqnarray}
and in the BPS limit when the Higgs potential vanishes, 
\begin{eqnarray}
{\cal E} &=& \mp\int\partial_i(B^a_i\Phi^a)~d^3 x + \int\frac{1}{2}(B^a_i \pm D_i\Phi^a)^2~d^3 x\nonumber\\
&=& \mp\int\partial_i(B^a_i\Phi^a)~d^3 x = 4\pi {\cal M}\xi,
\label{eq.6}
\end{eqnarray}
where ${\cal M}$ is the ``topological charge". 
The tensor identified with the electromagnetic field, as proposed by 't Hooft \cite{kn:2} is
\begin{eqnarray}
F_{\mu\nu} &=& \hat{\Phi}^a F^a_{\mu\nu} - \epsilon^{abc}\hat{\Phi}^{a}D_{\mu}\hat{\Phi}^{b}D_{\nu}\hat{\Phi}^c,\nonumber\\
	&=& \partial_{\mu}A_\nu - \partial_{\nu}A_\mu - \epsilon^{abc}\hat{\Phi}^{a}\partial_{\mu}\hat{\Phi}^{b}\partial_{\nu}\hat{\Phi}^c,
\label{eq.7}
\end{eqnarray}

\noindent where $A_\mu = \hat{\Phi}^{a}A^a_\mu$, the Higgs unit vector, $\hat{\Phi}^a = \Phi^a/|\Phi|$, and the Higgs field magnitude $|\Phi| = \sqrt{\Phi^{a}\Phi^{a}}$. 
The Abelian electric field is $E_i = F_{0i}$, and the Abelian magnetic field is $B_i = -\frac{1}{2}\epsilon_{ijk}F_{jk}$. 
We would also like to write the Abelian 't Hooft electromagnetic field as
\begin{equation}
F_{\mu\nu} = G_{\mu\nu} + H_{\mu\nu},
\label{eq.8}
\end{equation}

\noindent where 
\begin{eqnarray}
G_{\mu\nu} &=&  \partial_{\mu}A_\nu - \partial_{\nu}A_\mu, \nonumber\\
H_{\mu\nu} &=& - \epsilon^{abc}\hat{\Phi}^{a}\partial_{\mu}\hat{\Phi}^{b}\partial_{\nu}\hat{\Phi}^c,
\label{eq.9}
\end{eqnarray}

\noindent which we refer to as the gauge part and the Higgs part of the 't Hooft electomagnetic field respectively. 

The topological magnetic current \cite{kn:6} is defined to be 
\begin{equation}
k_\mu = \frac{1}{8\pi}~\epsilon_{\mu\nu\rho\sigma}~\epsilon_{abc}~\partial^{\nu}\hat{\Phi}^{a}~\partial^{\rho}\hat{\Phi}^{b}~\partial^{\sigma}\hat{\Phi}^c,
\label{eq.10}
\end{equation}

\noindent which is also the topological current density of the system and the corresponding conserved topological magnetic charge is
\begin{eqnarray}
{\cal M} & = & \int d^{3}x~k_0 = \frac{1}{8\pi}\int \epsilon_{ijk}\epsilon^{abc}\partial_{i}\left(\hat{\Phi}^{a}\partial_{j}\hat{\Phi}^{b}\partial_{k}\hat{\Phi}^{c}\right)d^{3}x \nonumber\\
& = & \frac{1}{8\pi}\oint d^{2}\sigma_{i}\left(\epsilon_{ijk}\epsilon^{abc}\hat{\Phi}^{a}\partial_{j}\hat{\Phi}^{b}\partial_{k}\hat{\Phi}^{c}\right)\nonumber\\
& = & \frac{1}{4\pi} \oint d^{2}\sigma_{i}~B_i. 
\label{eq.11}
\end{eqnarray}

\noindent As mentioned by Arafune et al. \cite{kn:7}, the magnetic charge ${\cal M}$ is the total magnetic charge of the system if and only if the gauge field is nonsingular. If the gauge field is singular and carries Dirac string monopoles, then the total magnetic charge of the system is the sum of the Dirac string monopoles and the monopoles carry by the Higgs field.


\section{The Exact Generalized Asymptotic Solutions}
The magnetic ansatz of Ref.\cite{kn:5} possesses internal space unit vectors with $\theta$-winding number $m$ and $\phi$-winding number $n$ greater than one,
\begin{eqnarray}
A_i^a &=&  - \frac{1}{r}\psi_1(r, \theta) \hat{u}^{a}_\phi\hat{\theta}_i + \frac{1}{r}\psi_2(r, \theta)\hat{u}^{a}_\theta\hat{\phi}_i
+ \frac{1}{r}R_1(r, \theta)\hat{u}^{a}_\phi\hat{r}_i - \frac{1}{r}R_2(r, \theta)\hat{u}^{a}_r\hat{\phi}_i, \nonumber\\
A^a_0 &=& 0, ~~\Phi^a = \Phi_1(r, \theta)~\hat{u}^a_r + \Phi_2(r, \theta)\hat{u}^a_\theta,
\label{eq.12}
\end{eqnarray}
\noindent where the spatial spherical coordinate orthonormal unit vectors are 
\begin{eqnarray}
\hat{r}_i &=& \sin\theta ~\cos \phi ~\delta_{i1} + \sin\theta ~\sin \phi ~\delta_{i2} + \cos\theta~\delta_{i3}, \nonumber\\
\hat{\theta}_i &=& \cos\theta ~\cos \phi ~\delta_{i1} + \cos\theta ~\sin \phi ~\delta_{i2} - \sin\theta ~\delta_{i3}, \nonumber\\
\hat{\phi}_i &=& -\sin \phi ~\delta_{i1} + \cos \phi ~\delta_{i2},
\label{eq.13}
\end{eqnarray}
and the isospin coordinate orthonormal unit vectors are 
\begin{eqnarray}
\hat{u}_r^a &=& \sin m\theta ~\cos n\phi ~\delta_{1}^a + \sin m\theta ~\sin n\phi ~\delta_{2}^a + \cos m\theta~\delta_{3}^a,\nonumber\\
\hat{u}_\theta^a &=& \cos m\theta ~\cos n\phi ~\delta_{1}^a + \cos m\theta ~\sin n\phi ~\delta_{2}^a - \sin m\theta ~\delta_{3}^a,\nonumber\\
\hat{u}_\phi^a &=& -\sin n\phi ~\delta_{1}^a + \cos n\phi ~\delta_{2}^a; ~~~\mbox{where}~~m\geq 1, ~~n\geq 1.
\label{eq.14}
\end{eqnarray}

\noindent In Ref.\cite{kn:5}, the MAP solutions and the MAC solutions corresponds to MA-MA-.....-MA pairs and M-A-M-A-.......-M-A-M chain respectively lying on the $z$-axis. Hence the MAP solutions possess zero net magnetic charge whereas the MAC solutions have a net magnetic charge of $n=1, 2$. 
When $m=2, 4, 6, ....$, the solutions are the 1-MAP, 2-MAP, 3-MAP, ...., configurations respectively and
when $m=1, 3, 5, ....$, the solutions are the M, M-A-M, M-A-M-A-M, ......, monopole chains respectively.  
Kleihaus et al. \cite{kn:5} have shown that these MAP and MAC solutions can have a single pole charge of $n=1, 2$. When the $\phi$-winding number $n=3$ and above, vortex rings are formed when $m\geq 2$.

By using the relationship \cite{kn:8}
\begin{eqnarray}
\hat{u}^a_r &=& \cos(m-1)\theta ~\hat{n}_r^a + \sin(m-1)\theta ~\hat{n}_\theta^a, \nonumber\\
\hat{u}^a_\theta &=& -\sin(m-1)\theta ~\hat{n}_r^a + \cos(m-1)\theta ~\hat{n}_\theta^a, ~~
\hat{u}^a_\phi = \hat{n}_\phi^a
\label{eq.15}
\end{eqnarray}
where $\hat{n}_r^a$, $\hat{n}_\theta^a$, and $\hat{n}_\phi^a$ are the unit vectors of Eq.(\ref{eq.14}) when $m=1$,
we can reduce the MAP and MAC solutions of Ref. \cite{kn:5} to the $\theta$-winding number, $m=1$, solutions. Hence the asymptotic solutions with $m=1$ at large $r$ is given by

\begin{eqnarray}
\psi_1&=& 1+p, ~\psi_2 = \frac{n(\sin\theta + \sin p\theta (a\cos\theta+b))}{\sin\theta}, ~
R_1 = 0, ~\nonumber\\
R_2 &=& \frac{n(\cos\theta - \cos p\theta(a\cos\theta+b))}{\sin\theta}, ~
\Phi_1 = \xi \cos p\theta, ~\Phi_2=\xi \sin p\theta,
\label{eq.16}
\end{eqnarray}
with $a=0$, $b=1$, and $p=1, 3, 5, ...$ in the MAP solution and $a=1$, $b=0$, and $p=0, 2, 4, ...$ in the MAC solution. Here we notice that the solution (\ref{eq.16}) can be generalized to 
\begin{eqnarray}
\psi_1&=& 1+\frac{\partial_\theta h(\infty,\theta)}{g(\infty,\theta)}, ~~~\psi_2 = \frac{n(\sin\theta + h(\infty,\theta)(a\cos\theta + b))}{\sin\theta}, ~~\nonumber\\
R_1 &=& 0, ~~~R_2 = \frac{n(\cos\theta - g(\infty,\theta)(a\cos\theta+b))}{\sin\theta}, ~~\nonumber\\
\Phi_1 &=& \xi ~g(\infty,\theta), ~~~\Phi_2=\xi ~h(\infty,\theta),
\label{eq.17}
\end{eqnarray}
where $g(r,\theta)^2+h(r,\theta)^2=1$ and $a$, $b$, $\xi$ are constants. In Ref.\cite{kn:8}, we notice that the solution can be generalized to include the Jacobi elliptic functions $E_A = E_A(u, k)$; $A=1,2$, as follow,
\begin{eqnarray} 
g(\infty,\theta) = E_1(u,k) ~~\mbox{and}~~ h(\infty,\theta) = E_2(u,k).
\label{eq.18}
\end{eqnarray}
Here $u=p \theta$ and $E_1(u, k)$ and $E_2(u, k)$ are a pair of non-singular Jacobi elliptic functions that satisfy the relation  $E_1(u, k)^2+E_2(u, k)^2=1$.

The three possible pairs of non-singular Jacobi elliptic solutions are 
\begin{eqnarray}
E_1(u,k) &=& cn(u,k) ~~\mbox{and}~~ E_2(u,k)= sn(u,k) 
\label{eq.19}\\
E_1(u,k) &=& dn(u,k) ~~\mbox{and}~~ E_2(u,k)= k~sn(u,k) 
\label{eq.20}\\
E_1(u,k) &=& \frac{cn(u,k)}{dn(u,k)} ~~\mbox{and}~~ E_2(u,k)= \frac{k^{\prime}sn(u,k)}{dn(u,k)},
\label{eq.21}
\end{eqnarray}
where $0\leq k\leq 1$ is the Jacobi elliptic parameter and $k^{\prime}=\sqrt{1-k^2}$. We then label the three sets of solutions (\ref{eq.19}) - (\ref{eq.21}) by calling them the Jacobi elliptic A (JEA), Jacobi elliptic B (JEB), and Jacobi elliptic C (JEC) solutions respectively. One should also notice that when $k=0$, the JEA and JEC solutions revert back to solution (\ref{eq.16}) and the JEB solution becomes the 't Hooft-Polyakov one-monopole solution when $a=1$, $b=0$, and $n=1$.

We have also notice that an even more general asymptotic solution exists when
\begin{eqnarray}
g(\infty,\theta) &=& \sin q\theta ~E_2(u,k) + \cos q\theta ~E_1(u,k),
\label{eq.22}\\
h(\infty,\theta) &=& \cos q\theta ~E_2(u,k) - \sin q\theta ~E_1(u,k),
\label{eq.23}
\end{eqnarray}
where $q$ is a constant.
When $E_A(u,k)$ is given by Eq.(\ref{eq.19}), we have the JEA type solution.
The gauge potentials and Higgs fields of the one-monopole JEA type solution at large $r$ is then given by
\begin{eqnarray}
A_i^a &=& -\frac{1}{r}\{1-q+p~dn(p\theta, k)\}\hat{\phi}^{a}\hat{\theta}_i \nonumber\\
&+& \frac{1}{r\sin\theta}\{\sin\theta + \{\cos q\theta ~sn(p\theta, k)-\sin q\theta ~cn(p\theta, k)\}\cos\theta\}\hat{\theta}^{a}\hat{\phi}_i \nonumber\\
&-& \frac{1}{r\sin\theta}\{\cos \theta - \{\sin q\theta ~sn(p\theta, k)+\cos q\theta ~cn(p\theta, k)\}\cos\theta\}\hat{r}^{a}\hat{\phi}_i, \nonumber\\
A^a_0 &=& 0, \nonumber\\
\Phi^a &=& \xi \{\sin q\theta ~sn(p\theta, k)+\cos q\theta ~cn(p\theta, k)\}~\hat{r}^a \nonumber\\
&+& \xi \{\cos q\theta ~sn(p\theta, k)-\sin q\theta ~cn(p\theta, k)\}\hat{\theta}^a.
\label{eq.24}
\end{eqnarray}
By using suitable choices for the solution constants $(p,q,k)$, the gauge potentials of Eq.(\ref{eq.24}) can be made non singular. In this paper we solved numerically for four one-monopole JEA type solutions when $k=k_1=0.9844325133$ or when the period of the Jacobi elliptic function A is $4\pi$,
\begin{eqnarray}
\int^{\frac{\pi}{2}}_0 \frac{d\theta}{\sqrt{1-k^2\sin^2\theta}}=\pi.
\label{eq.25}
\end{eqnarray}
Four axially symmetric one-monopole solutions are obtained when $(p,q,k) = (1,\frac{1}{2},k_1), (1,\frac{5}{2},k_1), (2,1,k_1),$ and $(2,3,k_1)$. When $(p,q,k) = (0,0,k)$, the solution is just the 't Hooft-Polyakov one-monopole. 
The solution (\ref{eq.24}) satisfied the temporal gauge condition, $A^a_0=0$, only. 
The non-Abelian magnetic fields and the Higgs field covariant derivatives are given respectively by 
\begin{eqnarray}
B^a_i &=& \frac{1}{r^2}(\dot{R_2}+R_2\cot\theta+\psi_1+\psi_2-\psi_1\psi_2)\hat{r}^a\hat{r}_i  \nonumber\\
&+& \frac{1}{r^2}(-\dot{\psi_2}+(\psi_1-\psi_2)\cot\theta+(1-\psi_1)R_2)\hat{\theta}^a\hat{r}_i \nonumber\\
&+& \frac{1}{r^2}(-rR^\prime_2+(1-\psi_2)R_1)\hat{r}^a\hat{\theta}_i \nonumber\\
&+& \frac{1}{r^2}(r\psi^\prime_2+(\cot\theta-R_2)R_1)\hat{\theta}^a\hat{\theta}_i + \frac{1}{r^2}(r\psi^\prime_1+\dot{R_1})\hat{\phi}^a\hat{\phi}_i  \nonumber\\
&=& B_i\hat{\Phi}^a 
\label{eq.26}\\
D_i\Phi^a &=&  (\Phi_1^\prime-\frac{1}{r}R_1\Phi_2)\hat{r}^a\hat{r}_i + (\Phi_2^\prime+\frac{1}{r}R_1\Phi_1)\hat{\theta}^a\hat{r}_i \nonumber\\
&+& \frac{1}{r}(\dot{\Phi}_1-(1-\psi_1)\Phi_2)\hat{r}^a\hat{\theta}_i + \frac{1}{r}(\dot{\Phi}_2+(1-\psi_1)\Phi_1)\hat{\theta}^a\hat{\theta}_i \nonumber\\
&+& \frac{1}{r}((1-\psi_2)\Phi_1+(\cot\theta-R_2)\Phi_2)\hat{\phi}^a\hat{\phi}_i\nonumber\\
&=&0,
\label{eq.27}
\end{eqnarray} 
From Eq.(\ref{eq.7}) - (\ref{eq.9}), the 't Hooft Abelian magnetic field $B_i$ is given by
\begin{eqnarray}
B_i &=& B^G_i + B^H_i, ~~~\nonumber\\
B^G_i &=& -\epsilon_{ijk}\partial_j A_k ~~\mbox{and}~~ B^H_i = \frac{1}{2}\epsilon_{ijk}\epsilon^{abc}\hat{\Phi}^a\partial_j\hat{\Phi}^b\partial_k\hat{\Phi}^c. 
\label{eq.28}
\end{eqnarray}
In Ref.\cite{kn:9}, it mentioned that $A_\mu$ is the Abelian component of $A^a_\mu$, and that it is the unrestricted part of the gauge potential $A^a_\mu$, that is, it is not restricted by the condition,
\begin{eqnarray}
D_\mu\hat{\Phi}^a = \partial_\mu\hat{\Phi}^a + \epsilon^{abc}A^b_\mu\hat{\Phi}^c=0. \nonumber
\end{eqnarray}
In fact the general large $r$ asymptotic solution (\ref{eq.17}) can be written as 
\begin{eqnarray}
A^a_\mu &=& A_\mu~\hat{\Phi}^a - \epsilon^{abc}\hat{\Phi}^b\partial_\mu \hat{\Phi}^c,\nonumber\\
\mbox{where} ~~~A_\mu &=& A~\hat{\phi}_\mu, ~~~~~\hat{\Phi}^a=g ~\hat{r}^a+h ~\hat{\theta}^a, \nonumber\\
A &=& \frac{1}{r}(\psi_2~h - R_2~g) = \frac{h\sin\theta -g\cos\theta +a\cos\theta +b}{r\sin\theta}.\nonumber
\end{eqnarray}
The other component of $A^a_\mu$ that is restricted is $- \epsilon^{abc}\hat{\Phi}^b\partial_\mu \hat{\Phi}^c$ and it is completely determined by the magnetic symmetry. Hence the unrestricted component of $A^a_\mu$ is referred to as the electric part and the restricted component of $A^a_\mu$ is referred to as the magnetic part of the gauge potential $A^a_\mu$ \cite{kn:9}. 

With some calculations, it can be shown that the magnetic part of $B^a_i$ which is also the Higgs part is
\begin{eqnarray}
B^H_i &=& -n\epsilon_{ijk}\partial_j(g\cos\theta - h\sin\theta) ~\partial_k \phi, 
\label{eq.29}
\end{eqnarray}
and the electric part which is also the gauge part is
\begin{eqnarray}
B^G_i &=& -n\epsilon_{ijk}\partial_j\left(\frac{\sin\theta}{n}(\psi_2 h -R_2 g)\right) ~\partial_k \phi,
\label{eq.30}
\end{eqnarray}
where $B^H_i$ is from the magnetic component of $A^a_\mu$ and $B^G_i$ is from the electric component of $A^a_\mu$.
Therefore the net magnetic field when $n=1$ is
\begin{eqnarray}
B_i &=&-n\epsilon_{ijk}\partial_j\left\{\cos\alpha(r,\theta)\right\} ~\partial_k \phi, \nonumber\\
\mbox{where} ~~\cos\alpha(r,\theta)&=&\sin\theta(g(\cot\theta-R_2)-h(1-\psi_2)). 
\label{eq.31}
\end{eqnarray}
A contour plot of $\cos\alpha(r,\theta) =$ constant at a fixed value of $\phi$ will give the magnetic field lines of the one-monopole solutions. In the limit when $r\rightarrow\infty$, $n=a=1$ and $b=0$,
\[B_i = -\epsilon_{ijk}\partial_j\left\{\cos\theta\right\} ~\partial_k \phi = \frac{\hat{r}_i}{r^2}\]
is just the magnetic field of a one-monopole. The Abelian energy density of the monopole is then given by
\begin{eqnarray}
\frac{1}{2}B_iB^i=\frac{1}{2r^2\sin^2\theta}\partial_j\cos\alpha~\partial^j\cos\alpha.
\label{eq.32}
\end{eqnarray}

\section{The Numerical One-Monopole Solutions}



\begin{figure}[tbh]
	\centering
		\includegraphics[scale=0.65]{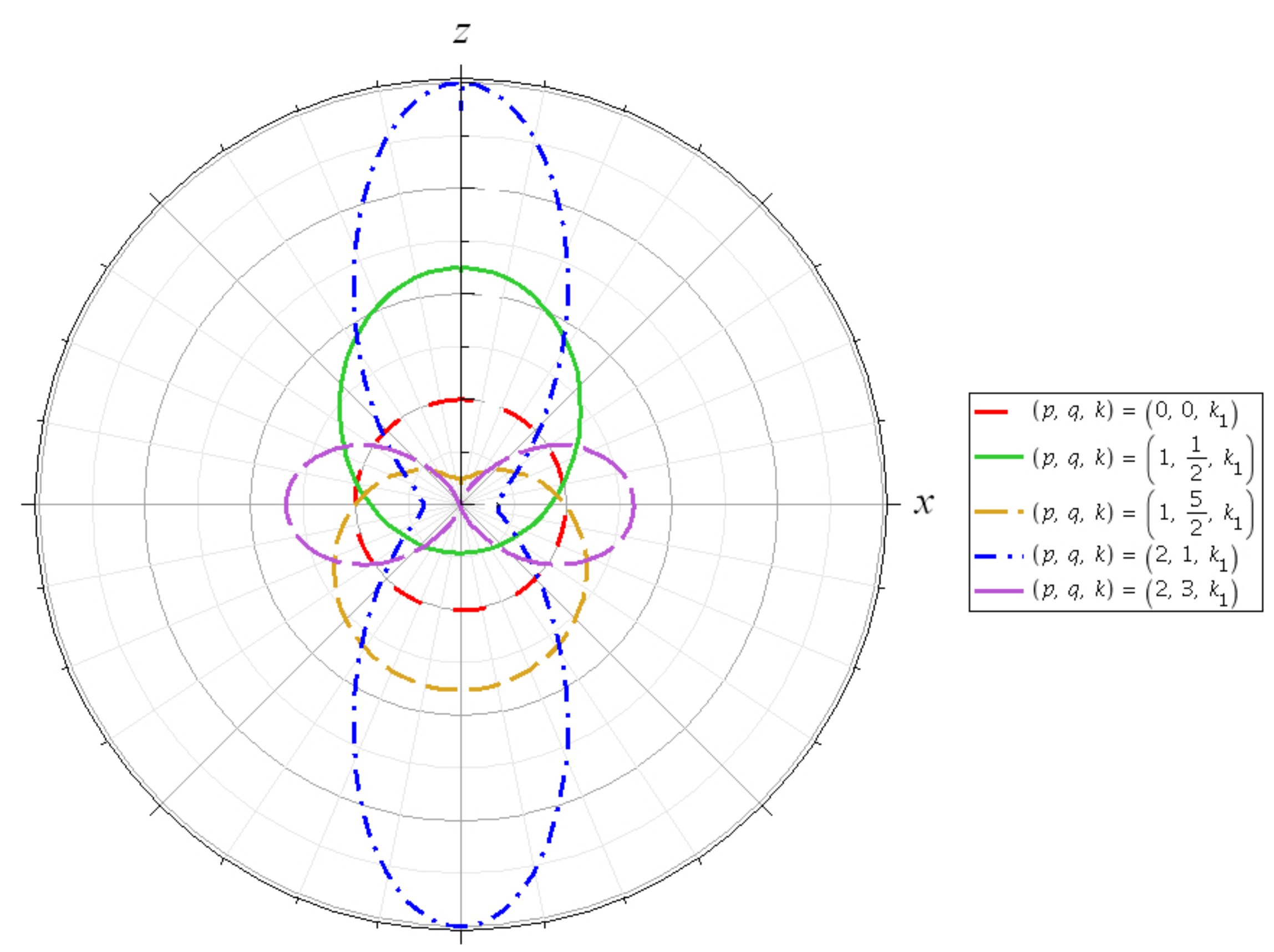}
	\caption{A polar plot of the magnitude of the Higgs magnetic field, $|r^2 B^H_i|$ versus $\theta$, at large $r$ for the 't Hooft-Polyakov monopole and the four new monopoles.}
	\label{fig.1}
\end{figure}



\begin{figure}[tbh]
	\centering
		\includegraphics[scale=0.5]{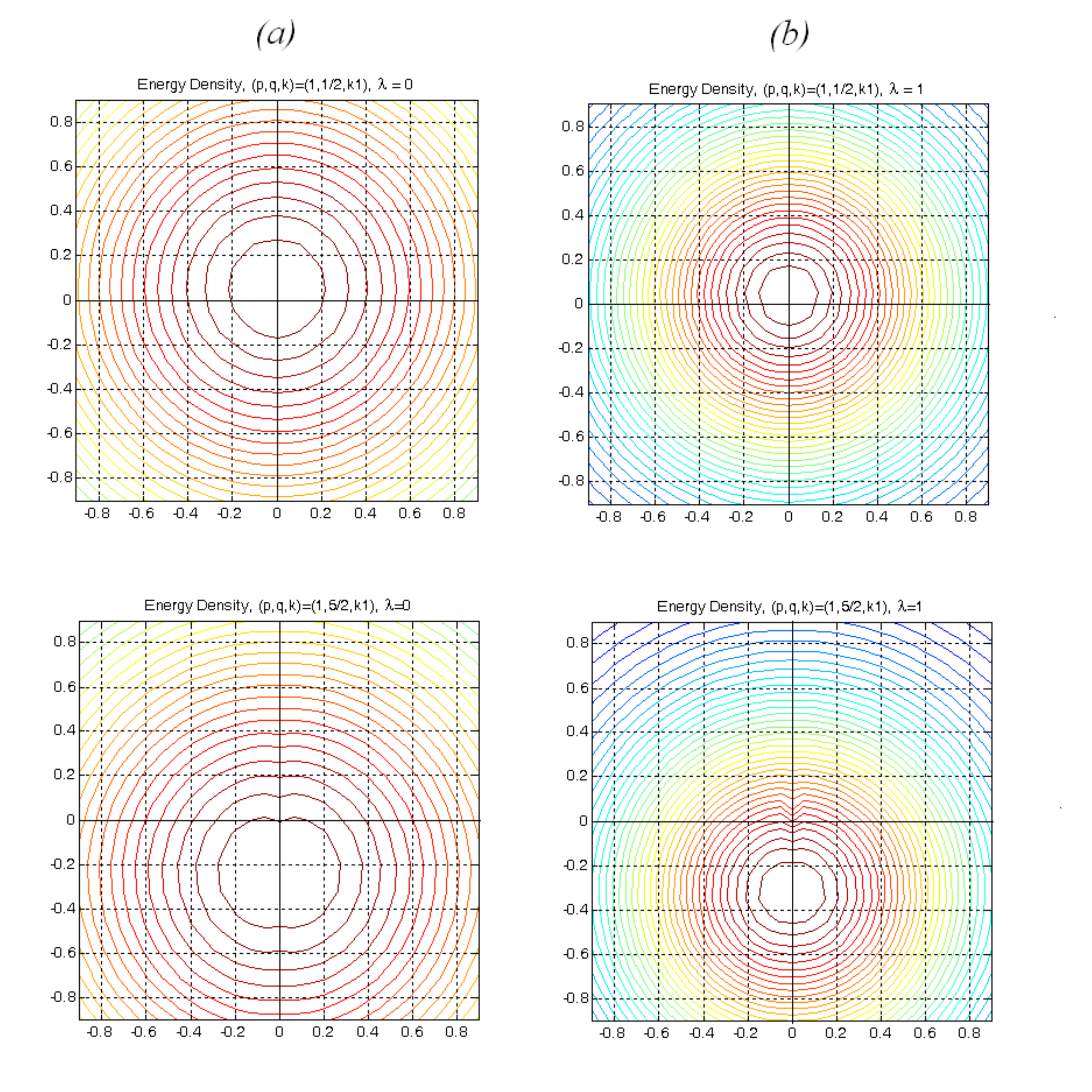}
	\caption{A polar contour plot of the energy density of the $(p,q,k)=(1,\frac{1}{2},k_1)$ one-monopole and the $(p,q,k)=(1,\frac{5}{2},k_1)$ one-monopole when (a) $\lambda=0$ and (b) when $\lambda=1$.}
	\label{fig.2}
\end{figure}


\begin{figure}[tbh]
	\centering
		\includegraphics[scale=0.5]{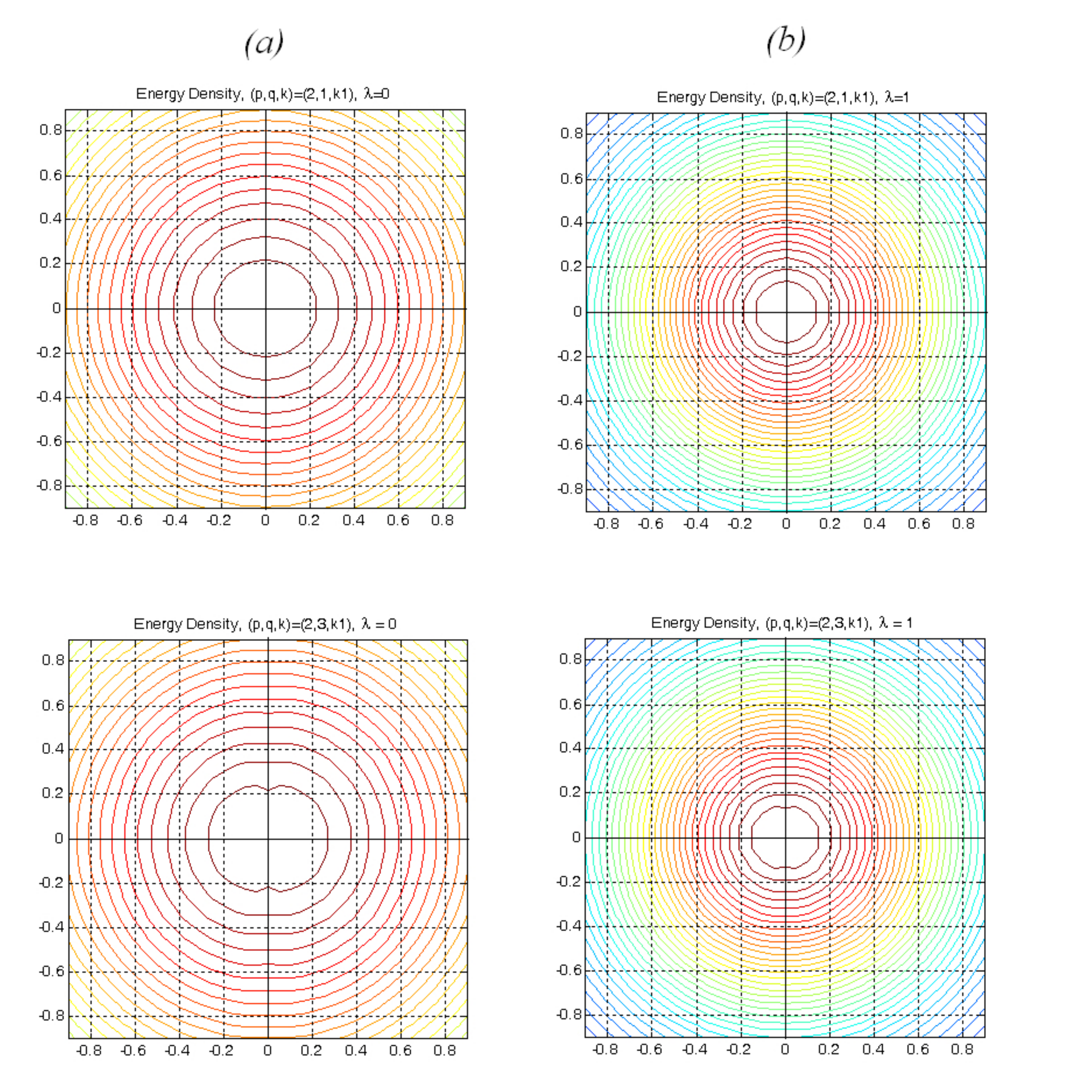}
	\caption{A polar contour plot of the energy density of the $(p,q,k)=(2,1,k_1)$ one-monopole and the $(p,q,k)=(2,3,k_1)$ one-monopole when (a) $\lambda=0$ and (b) when $\lambda=1$.}
	\label{fig.3}
\end{figure}


\begin{figure}[tbh]
	\centering
		\includegraphics[scale=0.56]{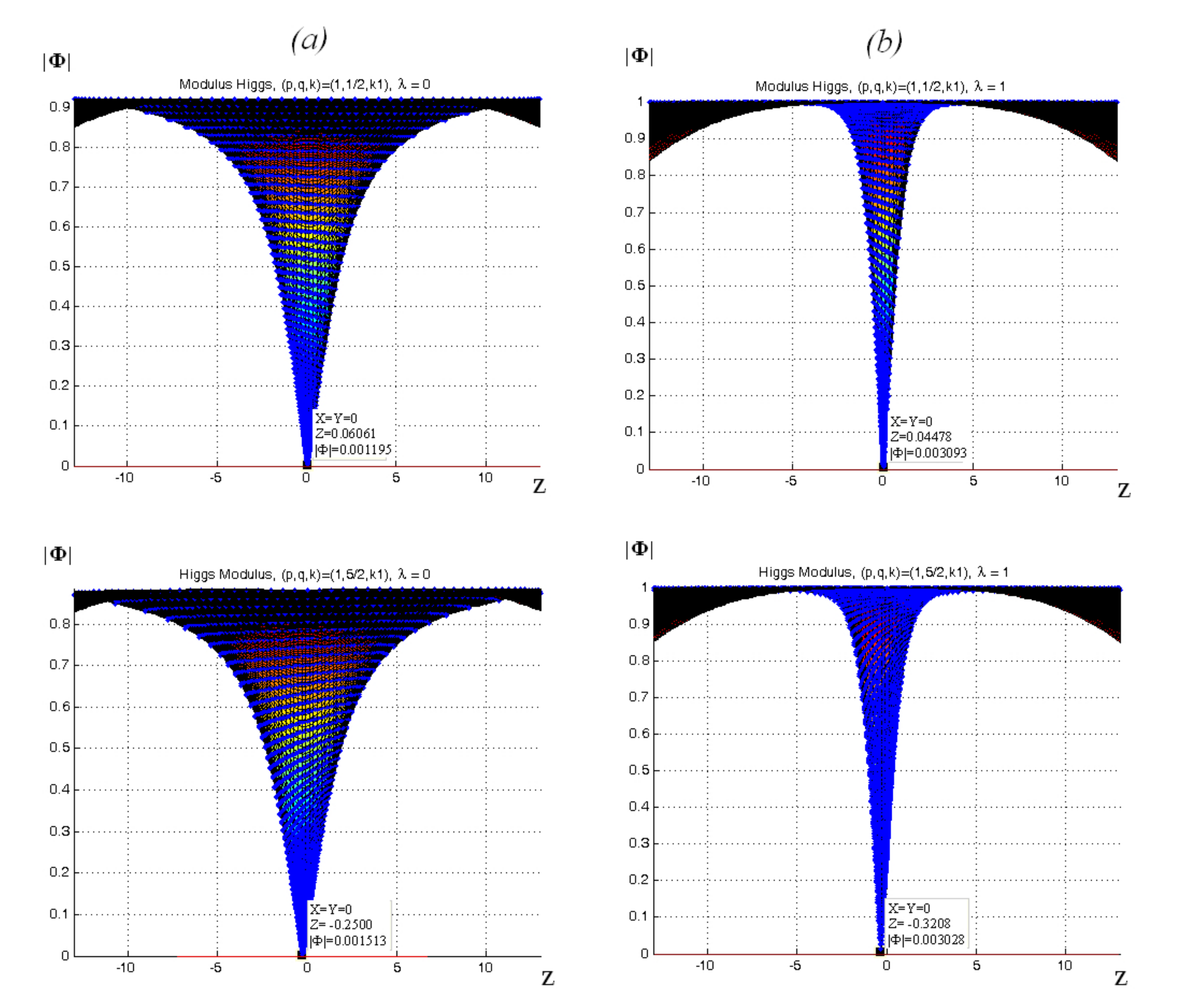}
	\caption{A 3D plot of the Higgs field modulus $|\Phi|$ of the $(p,q,k)=(1,\frac{1}{2},k_1)$ one-monopole and the $(p,q,k)=(1,\frac{5}{2},k_1)$ one-monopole when (a) $\lambda=0$ and (b) when $\lambda=1$.}
	\label{fig.4}
\end{figure}


\begin{figure}[tbh]
	\centering
		\includegraphics[scale=0.56]{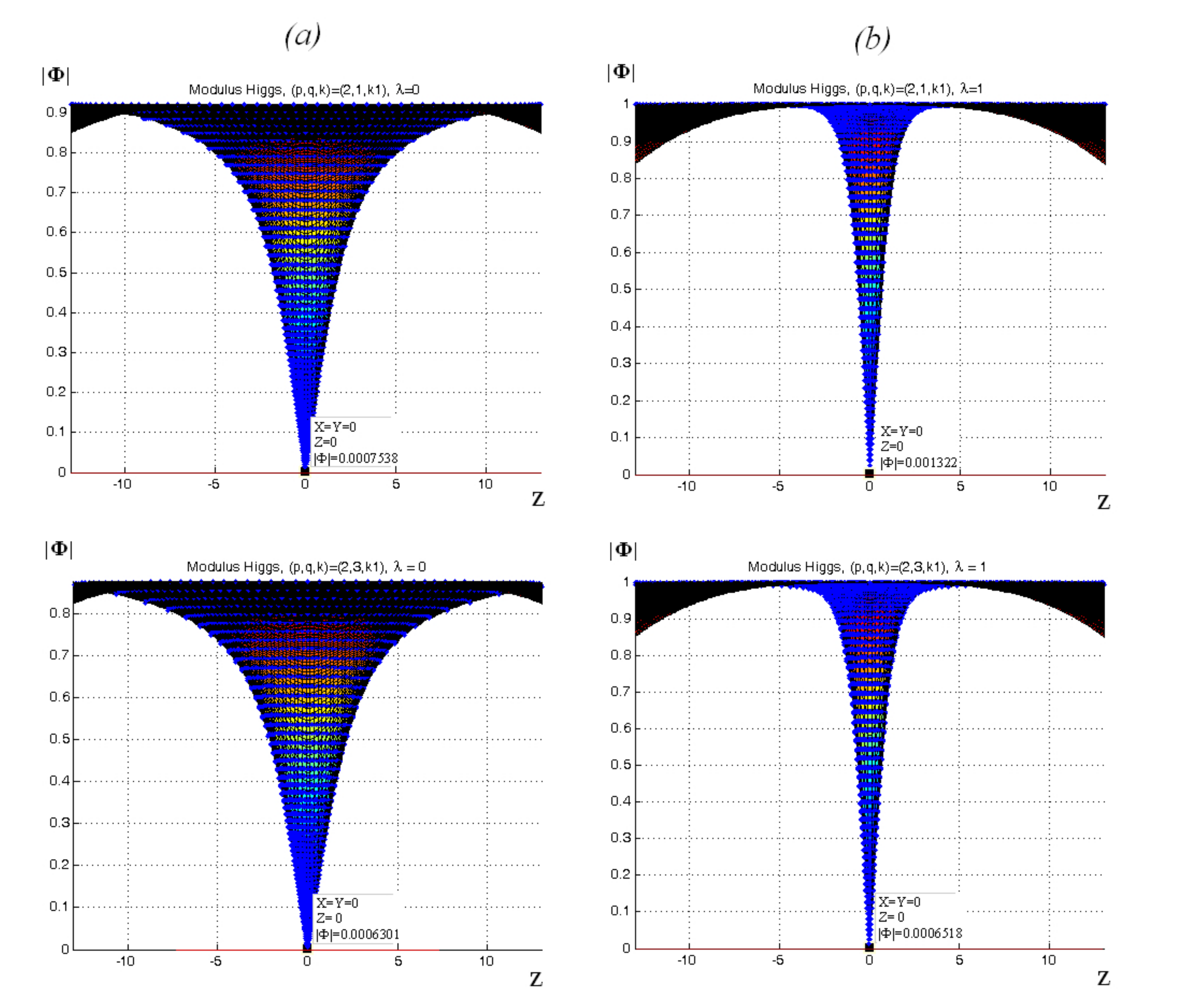}
	\caption{A 3D plot of the Higgs field modulus $|\Phi|$ of the $(p,q,k)=(2,1,k_1)$ one-monopole and the $(p,q,k)=(2,3,k_1)$ one-monopole when (a) $\lambda=0$ and (b) when $\lambda=1$.}
	\label{fig.5}
\end{figure}


The numerical calculations were performed using the Maple 12 and MatLab R2009a softwares. The second order equations of motion (\ref{eq.3}) are reduced to six partial differential equations with the ansatz (\ref{eq.12}) which are then transformed into a system of nonlinear equations using the finite difference approximation. This system of nonlinear equations are then discretized on a non-equidistant grid of size $70\times60$ covering the integration regions $0\leq \bar{x} \leq 1$ and $0\leq \theta \leq \pi$. Here $\bar{x}$ is the finite interval compactified coordinate given by $\bar{x}=\frac{r}{r+1}$. The partial derivative with respect to the radial coordinate is then replaced accordingly by
$\partial_r \rightarrow (1-\bar{x})^2 \partial_{\bar{x}}$ and ~~$\frac{\partial^2}{\partial r^2} \rightarrow (1-\bar{x})^4\frac{\partial^2}{\partial \bar{x}^2} - 2(1-\bar{x})^3\frac{\partial}{\partial \bar{x}}$. First of all, we used Maple to find the Jacobian sparsity pattern for the system of nonlinear equations.  After that we provide this information to Matlab to run the numerical computation. The system of nonlinear equations are then solved numerically using the trust-region-reflective algorithm by providing the solver with good initial guess.

The second order equations of motion Eq. (\ref{eq.3}) were solved when the winding numbers $m=1$ and $n=1$ and firstly with zero Higgs potential, that is $\lambda=\mu=0$ but with nonzero expectation value $\xi=1$ and finally with $\lambda=\mu=1$ and $\xi=1$. Hence the solutions obtained are non-BPS solutions with total energy given by 
\begin{eqnarray}
{\cal E} \geq 4\pi\xi M. 
\label{eq.33}
\end{eqnarray}

The boundary conditions used at small $r$ is given by
\begin{eqnarray}
\psi_1(0,\theta)=\psi_2(0,\theta)=R_1(0,\theta)=R_2(0,\theta)&=&0\nonumber\\ 
\Phi_1(0,\theta)=\xi_0\cos\theta, ~~\Phi_2(0,\theta)=-\xi_0\sin\theta \nonumber\\
\sin\theta ~\Phi_1(0,\theta) + \cos\theta ~\Phi_2(0,\theta) &=& 0\nonumber\\
\left.\frac{\partial}{\partial r}\left(\cos\theta ~\Phi_1(r,\theta) - \sin\theta ~\Phi_2(r,\theta)\right)\right|_{r=0} &=& 0
\label{eq.34}
\end{eqnarray}
and at large $r$ is given by Eq.(\ref{eq.24}). In order for the solution to be regular along the $z$-axis, the constraint imposed on the solution is 
\begin{eqnarray}
\left.R_1(r,~\theta)\right|_{\theta=0, \pi}=\left.R_2(r,~\theta)\right|_{\theta=0, \pi}=\left.\Phi_2(r,~\theta)\right|_{\theta=0, \pi}&=&0,\nonumber\\ 
\left.\frac{\partial}{\partial \theta}\psi_1(r,~\theta)\right|_{\theta=0, \pi}=\left.\frac{\partial}{\partial \theta}\psi_2(r,~\theta)\right|_{\theta=0, \pi}=\left.\frac{\partial}{\partial \theta}\Phi_1(r,~\theta)\right|_{\theta=0, \pi}&=&0,
\label{eq.35}
\end{eqnarray}
The above constraints are sufficient to ensure the regularity of the solution and the gauge fixing condition
\begin{eqnarray}
r\frac{\partial}{\partial r}R_1(r,\theta)-\frac{\partial}{\partial \theta}\psi_1(r,\theta)=0,
\label{eq.36}
\end{eqnarray}
is not imposed upon the numerical solutions here as Eq.(\ref{eq.36}) will restrict the one-monopole solutions to just one solution, that is the 't Hooft-Polyakov one-monopole solution. Similarly if the gauge fixing condition Eq.(\ref{eq.36}) is imposed on the MAP and MAC solutions, then these solutions are restricted to those discussed in Ref. \cite{kn:5}. We also note that the MAP and MAC solutions of Ref. \cite{kn:5} does not satisfied the radiation gauge condition $\partial^i A^a_i=0$ in general. It is only the 't Hooft-Polyakov one-monopole and the one-MAP solution that will satisfied the radiation gauge when the winding number $n=1$. All the other MAC and MAP solutions do not satisfied the radiation gauge condition.

The boundaries conditions imposed on the gauge potentials at large $r$, Eq.(\ref{eq.24}), small $r$, Eq.(\ref{eq.34}), and along the $z$-axis, Eq.(\ref{eq.35}), are sufficient to ensure regularity of the solutions. The ansatz (\ref{eq.12}) will automatically ensure that the 't Hooft Abelian gauge potential,
\begin{eqnarray}
A_i = A^a_i \hat{\Phi}^a = \frac{1}{r}(\psi_2(r,\theta) h(r,\theta) - R_2(r,\theta) g(r,\theta))\hat{\phi}_i,
\label{eq.37}
\end{eqnarray}
obeys the radiation gauge condition, $\partial^i A_i=0$.

By using suitable initial values, we obtained the new one-monopole numerical solutions when the the constants $(p,q,k)$ of the asymptotic solution (\ref{eq.24}) takes the following sets of values 
\begin{eqnarray}
(p,q,k) &=& (1,\frac{1}{2},k_1),~~ (1, \frac{5}{2},k_1),~~ (2,1,k_1),~ ~(2,3,k_1)\nonumber\\
k_1 &=& 0.9844325133. 
\label{eq.38}
\end{eqnarray}
These JEA one-monopole solutions are axially symmetric monopoles with Jacobi elliptic functions of period $4\pi$ in the asymptotic large $r$ solutions. Upon solving these four solutions using our numerical procedures, we plotted the polar plot of the magnitude of the Higgs magnetic field $|r^2 B_i^H|$ of these monopole solutions together with the 't Hooft-Polyakov monopole's $|r^2 B_i^H|$ versus $\theta$, at large $r$, Fig.\ref{fig.1}. We also calculate for the energy density, 
\begin{eqnarray}
ED(r,\theta)=\frac{1}{2}(B_{i}^{a}B_{i}^{a}+D_{i}\Phi^{a}D_{i}\Phi^{a}) + \frac{1}{4}\lambda(\Phi^a\Phi^a - \frac{\mu^2}{\lambda})^2,
\label{eq.39}
\end{eqnarray}
at small and intermediate $r$. Polar contour plots of the lines of same energy density versus $\theta$ are shown in Fig.\ref{fig.2} and Fig.\ref{fig.3} for the four new one-monopoles for values of $\lambda=0$ and $\lambda=1$. Hence from the polar contour plots of $ED(r,\theta)$ versus the angle $\theta$, we can conclude that these JEA type one-monopoles are axially symmetric about the $z$-axis. From these plots it is clearly seen that the magnitude of the Higgs magnetic field $|r^2 B_i^H|$ at infinity determines the shape of the one-monopole and hence a rough idea about the energy density distribution at small $r$.

Near to the origin as $r$ tends to zero, $B_i^G \rightarrow 0$ and $B_i\approx B_i^H$. From the polar plots of $|r^2 B_i^H|$ (Fig.\ref{fig.1}) and the energy density contour plots (Fig.\ref{fig.2} and Fig.\ref{fig.3}) of the four axially symmetric one-monopole, we are able to conclude that the $(1,\frac{1}{2}, k_1)$ one-monopole is stretched towards the positive $z$-axis \cite{kn:10}. The position of the monopole can be read from the zero of the Higgs field versus the $x$-$z$ plane 3D plot (Fig.\ref{fig.4}). When $\lambda = 0$, the monopole is located along the $z$-axis at $z=0.06$ and when $\lambda=1$, the monopole is at $z=0.04$.

The $(1,\frac{5}{2},k_1)$ one-monopole is however stretched in the opposite direction, that is, it is stretched along the negative $z$-axis. In this case the monopole is located along the $z$-axis at $z=-0.25$ when $\lambda=0$, and when $\lambda=1$ it is located at $z=-0.32$ (Fig.\ref{fig.4}). 

However the $(2,1,k_1)$ and $(2,3,k_1)$ one-monopoles are of different shapes and both are equally compressed along both the positive and negative $z$-axis and are hence symmetrical about the $z$-axis and the $x$-$y$ plane. Both these monopoles are located at the origin (Fig.\ref{fig.5}).



The total energy $\frac{{\cal E}}{4\pi\xi}$ of these one-monopoles are also calculated numerically. Our numerical calculation gives the values as shown in Table \ref{table:1} for 't Hooft-Polyakov one-monopole and the four axially symmetric one-monopoles for values of $\lambda=0$ and $\lambda=1$. Since the minimum total energy cannot be less than one, we conclude that all the five one-monopoles have similar total energy equal to one. The deviation from 1 of the numerically calculated values gives the magnitude of the numerical error involves in the calculation. As the complexity of the one-monopole gets higher, the numerical error gets bigger. 

When $\lambda=1$, the total energies of all the five monopoles are calculated to be 1.29 in units of ${4\pi\xi}$. Although all the monopoles seem to be having the same energies which is an invariant quantity for all gauge equivalent solutions, their energy densities do differ from one another as shown in Fig. 2 and 3. Since it is the energy density distribution in space that determines the shape of the monopole, we can conclude that the four new monopole are axially symmetric monopoles. 

\begin{table}[tbh]
\caption{The total energy in unit of $4\pi\xi$ of the 't Hooft-Polyakov one-monopole and the four axially symmetric one-monopoles as calculated numerically when (a) $\lambda=0$ and (b) when $\lambda=1$.}
\bigskip
\label{table:1}
\begin{tabular}{|c|c|c|c|c|c|} \hline
~&~&~&~&~&~\\ 
$(p,q,k)$ & ~$(0,0, k)$~ & ~$(1,\frac{1}{2}, k_1)$~ & ~$(1,\frac{5}{2}, k_1)$~ & ~$(2,1, k_1)$~ & ~$(2,3, k_1)$~ \\
~&~&~&~&~&~\\ \hline
(a) $\lambda=0$ & 0.9996& 0.9994& 1.0138& 0.9982& 1.0122 \\ 
~&~&~&~&~&~\\ \hline
(b) $\lambda=1$ & 1.2914& 1.2914& 1.2875& 1.2915& 1.2872 \\ 
~&~&~&~&~&~\\ \hline
\end{tabular}
\end{table} 

We also notice that the position of the monopole is moved when the parameter $p=1$, Table \ref{table:2}. The monopole is located on the positive side of the $z$-axis when $q=\frac{1}{2}$ and on the negative $z$-axis when $q=\frac{5}{2}$.

\begin{table}[tbh]
\caption{The position of the 't Hooft-Polyakov one-monopole and the four axially symmetric one-monopoles when (a) $\lambda=0$ and (b) when $\lambda=1$.}
\bigskip
\label{table:2}
\begin{tabular}{|c|c|c|c|c|c|} \hline
~&~&~&~&~&~\\ 
$(p,q,k)$ & ~$(0,0, k)$~ & ~$(1,\frac{1}{2}, k_1)$~ & ~$(1,\frac{5}{2}, k_1)$~ & ~$(2,1, k_1)$~ & ~$(2,3, k_1)$~ \\
~&~&~&~&~&~\\ \hline
(a) $\lambda=0$ & (0,0,0)& (0,0,0.06)& (0,0,-0.25)& (0,0,0)& (0,0,0) \\ 
~&~&~&~&~&~\\ \hline
(b) $\lambda=1$ & (0,0,0)& (0,0,0.04)& (0,0,-0.32)& (0,0,0)& (0,0,0) \\ 
~&~&~&~&~&~\\ \hline
\end{tabular}
\end{table} 

\section{Comments}
Although the total energy is equal to one for all the five monopoles when $\lambda=0$ and equal to 1.29 when $\lambda=1$, the energy density distribution varies from one monopole to the next and the four new monopoles only possess cylindrical symmetries about the $z$-axis. We would also like to take note that our value for the energy when $\lambda=1$ is quite close to the numerical value obtained by Bogomol'nyi and Marinov (1976), 1.238, and the value obtained by Bais and Primack (1976), 1.30 \cite{kn:2}. However the numerical value obtained by Kleihaus et al. (2004) \cite{kn:5} is 1.41 which is quite different from our value of 1.29.

We run the numerical solution of the 't Hooft-Polyakov one-monopole when $(p,q,k)=(0,0,k)$ with similar grid size of $70\times60$ as a pilot test for comparison with our results of the new one-monopoles. Hence we have shown that it is possible to construct a non radially symmetric one-monopole with finite energy. In this case, we have found axially symmetric one-monopole with finite energy.

The gauge potentials of the Jacobi elliptic generalized solutions do not satisfied the radiation gauge condition. However even without this constraint, we can only find certain discrete configurations of axially symmetric one-monopole solutions. There do not exist a parameter that can be continuously varied from the 't Hooft-Polyakov monopole to the four various axially symmetric one-monopoles. The solutions exist for only certain discrete values of $(p,q,k)$. The solution is not finite for other values of $(p,q,k)$. At the moment we are also looking for numerical one-monopole solutions that can be continuously distorted from the 't Hooft-Polyakov one-monopole by varying a parameter in the the asymptotic large $r$ exact solution. If God is willing, this work will be reported in the near future. 

From Table \ref{table:1}, we also notice that for a fixed value of $p$, the complexity of the solutions and hence the numerical error increased with the next higher value of $q$. 

The uses of these one-monopole solutions are still not clear and why only certain shapes and sizes are allowed in Nature is also not yet understood. Hence further work can be done in this direction.


\section*{Acknowledgements}
The authors would like to thank the Ministry of Science, Technology and Innovation (MOSTI) of Malaysia for the award of ScienceFund research grant (Project Number: 06-01-05-SF0266).

\newpage


\begin{thebibliography}{99}
\bibitem[1]{kn:1} G. 't Hooft, Nucl. Phy. {\bf B79},  276 (1974);
A.M. Polyakov, Sov. Phys. - JETP {\bf 41}, 988 (1975); Phys. Lett. {\bf B59},  82 (1975); JETP Lett. {\bf 20}, 194 (1974).

\bibitem[2]{kn:2} E.B. Bogomol'nyi and M.S. Marinov, Sov. J. Nucl. Phys. {\bf 23}, 355 (1976);
M.K. Prasad and C.M. Sommerfield, Phys. Rev. Lett. {\bf 35}, 760 (1975);
E.B. Bogomol'nyi, Sov. J. Nucl. Phys. {\bf 24}, 449 (1976); F.A. Bais and J.R. Primack, Phys. Rev. {\bf D13}, 819 (1976).

\bibitem[3]{kn:3} C. Rebbi and P. Rossi, Phys. Rev. {\bf D22}, 2010 (1980);
R.S. Ward, Commun. Math. Phys. {\bf 79}, 317 (1981);
P. Forgacs, Z. Horvarth and L. Palla, Phys. Lett. {\bf B99}, 232 (1981); 
Nucl. Phys. {\bf B192}, 141 (1981);
M.K. Prasad, Commun. Math. Phys. {\bf 80}, 137 (1981);  M.K. Prasad and P. Rossi, Phys. Rev. {\bf D24}, 2182 (1981);
Rosy Teh, Int. J. Mod. Phys. {\bf A16}, 3479 (2001); Rosy Teh and K.M. Wong, Int. J. Mod. Phys. {\bf A19}, 371 (2004).

\bibitem[4]{kn:4} E.J. Weinberg and A.H. Guth, Phys. Rev. {\bf D14}, 1660 (1976).

\bibitem[5]{kn:5} Bernard R\"{u}ber, Ph.D. thesis, University of Bonn, 1985; B. Kleihaus and J. Kunz, Phys. Rev. {\bf D61}, 025003 (2000);
B. Kleihaus, J. Kunz, and Y. Shnir, Phys. Lett. {\bf B570}, 237 (2003); Phys. Rev. {\bf D 68}, 101701 (2003); Phys. Rev. {\bf D 70}, 065010 (2004).

\bibitem[6]{kn:6} N.S. Manton, Nucl. Phys. (N.Y.) {\bf B126}, 525 (1977).
	
\bibitem[7]{kn:7} J. Arafune, P.G.O. Freund, and C.J. Goebel, J. Math. Phys. {\bf 16} 433 (1975).

\bibitem[8]{kn:8} Rosy Teh and K.M. Wong, FRONTIERS IN PHYSICS: 3rd International Meeting, Kuala Lumpur (Malaysia), 12-16 January 2009, edited by S.P. Chia, M.R. Muhammad, and K. Ratnavelu,  ISBN: 978-0-7354-0687-2, AIP Conference Proceedings Volume 1150, 101 (2009).

\bibitem[9]{kn:9} Y.M. Cho, Phys. Rev. {\bf D 21}, 1080 (1980).


\bibitem[10]{kn:10} K.M. Wong and Rosy Teh, FRONTIERS IN PHYSICS: 3rd International Meeting, Kuala Lumpur (Malaysia), 12-16 January 2009, edited by S.P. Chia, M.R. Muhammad, and K. Ratnavelu,  ISBN: 978-0-7354-0687-2, AIP Conference Proceedings Volume 1150, 420 (2009).


\end{thebibliography}
\end{document}